\documentstyle[12pt]{article}

\renewcommand{\theequation}{\thesection.\arabic{equation}}
\newcommand{\be}{\begin{equation}}  \newcommand{\ee}{\end{equation}}
\newcommand{\bear}{\begin{eqnarray}}
\newcommand{\eear}{\end{eqnarray}}
\newcommand{\ba}{\begin{array}}   \newcommand{\ea}{\end{array}}
\newcommand{\lae}{\begin{array}{c}\,\sim\vspace{-28pt}\\< \end{array}}
\newcommand{\gae}{\begin{array}{c}\,\sim\vspace{-28pt}\\> \end{array}}
\def\vbr{$\vphantom{\sqrt{F_e^i}}$}

\input prepictex
\input pictex
\input postpictex
\newdimen\tdim
\tdim=\unitlength
\def\bar{\overline}
\def\stpltsmbl{\setplotsymbol ({\small .})}
\def\tarrow{\arrow <8\tdim> [.3,.6]}
\def\moose#1#2#3{\tarrow from #1 to #2 \plot #2 #3 /}

\begin{document}

\pagestyle{empty}
\begin{titlepage}
\def\thepage {}    

\title{\bf Top Quark Seesaw Theory \\ [2mm]
of Electroweak Symmetry Breaking
\\ [1cm]}

\author{\bf R.~Sekhar Chivukula$^1$,
Bogdan A.~Dobrescu$^2$, \\ [2mm]
\bf Howard Georgi$^3$,
Christopher T.~Hill$^{2,4}$
\\ [6mm]
{\small {\it $^1$Physics Department, Boston University}}\\
{\small {\it Boston, Massachusetts 02215, USA }}\thanks{e-mail
 address: sekhar@bu.edu}
\\ [3mm]
{\small {\it $^2$Fermi National Accelerator Laboratory}}\\
{\small {\it P.O. Box 500, Batavia, Illinois 60510, USA}} \thanks{e-mail
 addresses: bdob@fnal.gov, hill@fnal.gov}
\\ [3mm]
{\small {\it $^3$Harvard University}}\\
{\small {\it Cambridge, Massachusetts 02138, USA }}\thanks{e-mail
 address: georgi@physics.harvard.edu}
\\ [3mm]
{\small {\it $^4$ The University of Chicago }}\\
{\small {\it Chicago, Illinois 60637, USA }}\\ }

\date{ }

\maketitle

\vspace*{-15.8cm}
\noindent

\begin{flushleft}
September 21, 1998 \\ [1mm]
hep-ph/9809470 \\ [1mm]
\end{flushleft}

\vspace*{-2.8cm}
\begin{flushright}
FERMILAB-Pub-98/284-T \\ [1mm]
BUHEP-98-25 \\ [1mm]
HUTP-98/A067 \\ [1mm]
EFI--98--37 \\ [1mm]
\end{flushright}

\vspace*{13.1cm}

\baselineskip=18pt

\begin{abstract}

  {\normalsize
We study electroweak symmetry breaking involving the seesaw mechanism of
quark condensation. These models produce a composite 
Higgs boson involving the left-handed top quark, yet the top mass
arises naturally  at the observed scale.
We describe a schematic model which illustrates the general
dynamical ideas. 
We also consider a  generic low-energy effective theory which includes
 several composite scalars,
and we use the effective potential formalism to compute their spectrum.
We develop  a more
detailed model in which certain features of the schematic model
are replaced by additional dynamics. } 
\end{abstract}

\vfill
\end{titlepage}

\baselineskip=18pt
\renewcommand{\arraystretch}{1.5}
\pagestyle{plain}
\setcounter{page}{1}


\section{Introduction}
\setcounter{equation}{0}

The Higgs doublet of the standard model, used to break the electroweak
symmetry and generate all observed quark, lepton and gauge boson masses,
does not have to be a fundamental field.  In fact, the fermions
observed so far have the appropriate quantum numbers to provide the
constituents of a composite Higgs field.  Therefore, it is interesting to
consider the existence of some new, non-confining strong interactions which
bind the quarks and/or leptons within a composite Higgs field, giving rise
to a condensate (associated with a Higgs VEV) and to 
Higgs-Yukawa couplings.  

Due to its large mass, the top quark
is a natural candidate for providing a constituent to
a composite Higgs boson and an electroweak symmetry breaking (EWSB)
condensate \cite{bhl, cond}. However, the computation of the $W$ and $Z$
masses to leading order in $1/N_c$ ($N_c$ is the number of colors) shows
that the quark whose condensate gives the bulk of EWSB
must have a mass of order 0.6 TeV (in the absence of an excessively
fine-tuned version of the model in which the new strong dynamics is
placed at the GUT scale).  Such a heavy quark may, 
in principle, be part of a fourth generation,
but in that case one would have to worry about the proliferation of
weak-doublet fermions that contribute to the electroweak radiative parameter
$S$, and the top would not be directly involved in the EWSB mechanism.

In a previous letter \cite{dhseesaw} two of us introduced the idea of a
dynamical top quark seesaw mechanism.  Here the EWSB occurs via the
condensation of the left-handed top quark, $t_L$, with a new, right-handed
weak-singlet quark, which we refer to as a $\chi$-quark. The
$\chi_R$ quark has hypercharge $Y= 4/3 $ and thus is indistinguishable from the
right-handed top, $t_R$.  The dynamics which
leads to this condensate is essentially topcolor \cite{topcolor,tc2}. The 
fermionic mass scale of this weak-isospin $I=1/2$ condensate is large, of order
0.6 TeV. This corresponds to the formation of a dynamical boundstate
weak-doublet Higgs field, $\sim (\overline{\chi_R} t_L, \overline{\chi_R} b_L)$.
To leading order in $1/N_c$ this
yields a VEV for the Higgs boson of the 
appropriate electroweak scale, $v/\sqrt{2} \approx 175$
GeV.  However, the model also incorporates a new left-handed 
weak-singlet $\chi$-quark, with $Y= 4/3$.
The $\chi$-quarks condense amongst themselves through additional new dynamics
at still larger mass scales. Moreover, the left-handed $\chi$-quark
has a weak-singlet condensate with the right-handed top
quark. There is {\em ab initio} no direct left-handed top
condensate with the right-handed anti-top in this scheme
(or else this condensate is highly suppressed). 

Upon diagonalization of the fermionic mass matrix this admits
a conventional seesaw mechanism, yielding the physical top quark mass as an
eigenvalue that is less than the $600$ GeV matrix element. Thus, the top
quark mass can be adjusted naturally to its experimental value. The
diagonalization of the fermionic mass matrix in no way affects the fact
that the model has a composite Higgs doublet, with a VEV of 
$v/\sqrt{2} \approx 175$ GeV.
The mechanism incorporates $t_L$, which provides the source of
the weak $I=1/2$ quantum number of the composite Higgs boson, and thus the
origin of the EWSB vacuum condensate. Topcolor 
and any additional strong dynamics is
occurring at a multi-TeV scale, and the observed top quark mass arises
naturally, being suppressed by a ratio of $\sim $TeV scales.
Indeed, if a mechanism like this operates in nature, then
we have already observed the key $I=1/2$ element of EWSB at the Tevatron !  

There are several attractive features of this
mechanism.  First, while there are the
additional $\chi$ quarks involved in the strong dynamics, 
{\em these do not carry
 weak-isospin quantum numbers}. This is a remarkable advantage from the
point of view of model building.  The counting constraints of
technicolor, e.g., on the number of techniquarks from the
$S$ parameter, are essentially irrelevant for us, since we have only a
top quark condensate in the EWSB channels.  The constraints on custodial
symmetry violation, i.e., the value of the $\Delta \rho$ or
equivalently, $T$ parameter, are easily satisfied, being principally the
usual $m_t$ contribution, plus corrections suppressed by the
seesaw mechanism \cite{dhseesaw}.

Second, the models make a robust prediction about the nature of the
electroweak condensate: the left--handed top quark is unambiguously
identified as the electroweak--gauged condensate fermion.  The scheme
demands the presence of some kind of topcolor interactions, new strong
interactions associated with the formation of the top condensate. This
implies that QCD itself will change character at the multi-TeV scale
as it is embedded into the larger topcolor containing gauge group.
However, beyond the $I=1/2$ component of the EWSB, the remainder of the
structure, e.g., the $\chi$-quarks and the additional strong forces
which they feel, is somewhat arbitrary at this point.  

Third, the scheme implies that in the absence of the seesaw, the top
quark would have a larger mass, of order $600$ GeV. This in turn leads
to a relaxation of the constraints on the masses of topcolor colorons and
any additional heavy gauge bosons, permitting the full topcolor structure to
be moved to somewhat higher mass scales.  This gives more model-building
elbow room, and may reflect the reality of new strong dynamics.

We believe the top quark seesaw is a significant new idea in dynamical
models of EWSB and opens up a large range of new model building
possibilities. For that reason we will give a fairly detailed
discussion of the seesaw mechanism in this paper.  

We begin in Section 2 with the presentation
of a schematic model. Here the electroweak condensate involving
$\overline{t_L}$ and $\chi_R$ is driven by topcolor interactions, but the 
weak-singlet
condensates involving $\chi_{L,R}$ and $t_R$ 
are simply mass terms that we implement by hand. This
naturally separates the problem of EWSB from the weak-singlet physics in the 
$\chi_{L,R}$ and $t_R$ sector, which is the key advantage of the
seesaw mechanism. We derive the effective Lagrangian for
the dynamical Higgs and its interactions with matter using
the renormalization group approach in the large-$N_c$ fermion-bubble 
approximation.  The schematic model shows the emergence of the
Higgs boundstate and the formation of the $\overline{\chi_R}t_L$
condensate.  The schematic model provides a point of
departure for the construction of more elaborate models, and
the problem of generating light fermion Higgs-Yukawa couplings,
which we will not address in detail. We will briefly summarize options
for addressing the problem of the $b$-quark mass in the schematic model. 
The Higgs boson mass is large in the schematic
model, given by $2 m_{t\chi} \sim {\cal{O}}(1\ {\rm TeV}) $ 
in the large-$N_c$ fermion-bubble approximation, essentially saturating
the unitarity bound of the standard model \cite{lqt}.

In Section 3 we proceed with a more ambitious attempt to include
additional interactions amongst the minimal set of fermions
of the schematic model. 
This is a somewhat general construction, and it leads to additional
composite scalars, and new effects. 
We give a full effective potential analysis of this
scheme. Weak-singlet mass terms are still required as
in the schematic model to trigger the desired
tilting of the vacuum, though they can now be much smaller than in the
schematic model since the  additional strong, yet subcritical,
interactions can amplify the effects of these mass terms \cite{appel}. 
These  interactions push the potential close to a second order 
(or weakly first-order) phase 
transition and thus, the Higgs boson can be as light as $\sim 100$ GeV. 
This requires a partial degeneracy between
the weak-doublet and weak-singlet composite scalars. In the decoupling
limit the more general theory resembles the standard model with a light
Higgs boson.  

In Section 4, we describe a class of models incorporating the top quark
seesaw mechanism in which topcolor symmetry breaking is dynamically
generated. These models can replace the explicit
weak-singlet mass terms with additional
dynamics, in analogy to extended technicolor.
These models also allow in principle for the generation of masses of the
light quarks and accommodate intergenerational mixing. While these models
do not yet provide a complete explanation of flavor 
symmetry breaking, we regard them as an existence proof and a guide to future
theoretical investigations.

Section 5 summarizes our conclusions. In Appendix A we apply the effective 
potential formalism of Section 3 to the top quark seesaw model of 
ref.~\cite{dhseesaw}. In Appendix B we prove that the coupled gap equations 
used in ref.~\cite{dhseesaw} are equivalent with the stationarity conditions
of the effective potential derived in Section 3.


\section{A Schematic Model}

\setcounter{equation}{0}

In the present Section we will study a schematic model of the top quark
seesaw. This model will be a minimal version of the top seesaw and is
intended primarily to exhibit the essential physics. The schematic model
contains the elements of the third generation, the left-handed top-bottom
doublet, $\psi_{L}= (t_{L},b_{L})$, the right-handed top quark, $t_{R}$,
(we will postpone discussing the right-handed $b$-quark and associated fields
for the moment; indeed, the present model will not be anomaly free without
the inclusion of $b_R$ and associated fields, so we return to
consider it
below). We further introduce two weak-singlet fermions, $\chi _{R}$ and 
$\chi_{L},$ each having the quantum numbers of $t_{R}.$ The schematic
model exhibits the dynamical formation, via topcolor, of the Higgs doublet
as a composite field of the form:
\begin{equation}
\varphi = \left(
\begin{array}{c}
\overline{\chi _{R}}\,t_{L} \\
\overline{\chi _{R}}\,b_{L}
\end{array} \right) ~.
\end{equation}

We proceed by introducing an embedding of QCD into the gauge groups 
$SU(3)_{1}\times SU(3)_{2}$, with coupling constants $h_{1}$ and $h_{2}$
respectively. These symmetry groups are broken down to $SU(3)_{QCD}$ at a
high mass scale ${\cal V}$. The assignment of the elementary fermions to
representations under the full set of gauge groups $SU(3)_{1}\times
SU(3)_{2}\times SU(2)_{W}\times U(1)_{Y}$ is as follows:
\begin{equation}
\psi_{L} : \, ({\bf 3,1,2,} \, +1/3) \;\;\; ,  \;\;\
\chi_{R} : \, ({\bf 3,1,1,} \, +4/3) \;\;\; ,  \;\;\
t_{R}, \chi_{L} : \, ({\bf 1,3,1,} \, +4/3) ~.
\end{equation}
This set of fermions is incomplete: the representation specified has
$[SU(3)_1]^3$, $[SU(3)_2]^3$, and $U(1)_Y [SU(3)_{1,2}]^2$ gauge anomalies.
These anomalies will be canceled by fermions associated with either the
dynamical breaking of $SU(3)_1 \times SU(3)_2$, or with producing the
$b$-quark mass (a specific example of the latter case is given at the
end of this section). The dynamics of EWSB and
top-quark mass generation will not depend on the details of these
additional fermions.

We further introduce a scalar field, $\Phi$, transforming as
$({\bf \overline{3},3,1,} \, 0)$,  with negative $M_{\Phi }^{2}$ and an associated
quartic potential such that $\Phi $ develops a diagonal VEV,
\begin{equation}
\langle \Phi^i_j \rangle = {\cal V}\delta_{j}^{i}\,,
\end{equation}
and topcolor is broken to QCD:
\begin{equation}
SU(3)_{1}\times
SU(3)_{2}\longrightarrow SU(3)_{QCD} ~,
\end{equation}
yielding massless gluons and an octet of degenerate colorons with mass $M$ given by:
\begin{equation}
M^{2}= (h_{1}^{2}+h_{2}^{2})\,{\cal V}^{2}\,.
\end{equation}
In more complete models this symmetry breaking may arise dynamically,
but we describe it in terms of a VEV of a fundamental scalar field in
the present model for the sake of simplicity.

We now introduce a Yukawa coupling of the fermions $\chi_{L,R}$ to
$\Phi $ of the form:
\begin{equation}
-\xi\,\overline{\chi _{R}}\,\Phi\, \chi _{L}+{\rm h.c.}
\longrightarrow -\mu_{\chi \chi }\,
\overline{\chi }\chi
\end{equation}
We emphasize that this is an electroweak singlet mass term.
In this scheme $\xi$ is a perturbative coupling constant so ${\cal V} \gg \mu_{\chi \chi }$.
Finally, since both $t_{R}$ and $\chi _{L}$ carry identical topcolor and
$U(1)_{Y}$ quantum numbers we are free to include an explicit mass term,
also an electroweak singlet, of
the form:
\begin{equation}
-\mu_{\chi t}\,\overline{\chi _{L}}\,t_{R}+{\rm h.c.}
\end{equation}
The mass terms of $\overline{\chi _{L}}\,\chi _{R}$ and 
$\overline{\chi_{L}}\,t_{R}$ may arise dynamically in subsequent schemes,
and are
introduced by hand into the schematic model for purposes of illustration.
With these terms, the Lagrangian of the model at scales below the
coloron mass is $SU(3)_C\times SU(2)_W\times U(1)$ invariant and
becomes:
\begin{equation}
{\cal {L}}_{0}= {\cal {L}}_{\rm kinetic}-(\mu_{\chi \chi }\,\overline{\chi _{L}}
\,\chi_{R}
+\mu_{\chi t}\,\overline{\chi _{L}}\,t_{R}+{\rm h.c.})+{\cal {L}}_{\rm int}
\label{njll}
\end{equation}
${\cal {L}}_{\rm int}$ contains the residual topcolor interactions from the
exchange of the massive colorons:
\begin{equation}
{\cal {L}}_{\rm int}=  -\frac{g^{2}_{\rm tc}}{M^{2}}\left( \overline{\psi _{L}}\,
\gamma^{\mu }\frac{\lambda ^{A}}{2}\psi _{L}\right)
\left( \overline{\chi _{R}}\,
\gamma_{\mu }\frac{\lambda ^{A}}{2}\chi _{R}\right) +LL+RR\; ,
\label{njl0}
\end{equation}
where $LL$ $(RR)$ refers to left-handed (right-handed) current-current
interactions, and $g_{tc}$ is the topcolor gauge coupling.
Since the topcolor interactions are strongly coupled, forming 
boundstates, higher dimensional operators might have a significant 
effect on the low energy theory. However, if the full topcolor dynamics 
induces chiral symmetry breaking through a second order (or weakly first order)
phase transition, then one can 
analyze the theory using the fundamental degrees of freedom,
namely the quarks, at scales significantly lower than the topcolor scale. 
We will assume that this is the case, which implies that 
the effects of the higher dimensional operators are suppressed by powers of the
topcolor scale, and it is sufficient to keep in the low energy theory 
only the effects of the operators shown in eq.~(\ref{njl0}).
Furthermore, the $LL$ and $RR$ interactions do not affect the low-energy
effective potential in the large $N_c$ limit \cite{cg1}, so we will ignore them 
(one should keep in mind that these interactions may have other effects,
such as contributions to the custodial symmetry violation parameter $T$ 
\cite{cdt,cg1}, but these effects are negligible 
if the topcolor scale is in the multi-TeV range).

To leading order in $1/N_c$, the $LR$ interaction in (\ref{njl0}) can be 
rearranged into the following form:
\begin{equation}
{\cal {L}}_{\rm int}= \frac{g_{\rm tc}^{2}}{M^{2}}(\overline{\psi_{L}}\,
\chi_{R})\,(\overline{\chi_{R}}\,\psi _{L}) ~.
\label{njl1}
\end{equation}
This is the Nambu-Jona-Lasinio (NJL) interaction \cite{NJL}, which provides
the binding of the composite Higgs multiplet.
We will analyze the physics of (\ref{njll}) by using 
the coloron mass $M$ as a momentum space cut-off on the loop
integrals of the theory.

It is convenient to pass to a mass eigenbasis with the following
redefinitions:
\bear
\chi _{R}^{\prime } & = & \cos \theta \;\chi _{R}+\sin \theta\;
t_{R} ~,
\nonumber \\ [2mm]
\quad t_{R}^{\prime } & = & \cos \theta \;t_{R}-\sin \theta \;\chi _{R}  ~,
\eear
where:
\begin{equation}
\tan \theta = \frac{\mu_{\chi t}}{\mu_{\chi \chi }}
\end{equation}
In this basis, the NJL Lagrangian takes the form:
\bear
{\cal {L}}_{0} & = & {\cal {L}}_{\rm kinetic}-
\overline{M}\,\overline{\chi_{R}^{\prime }}\,\chi_{L}+{\rm h.c.}
\nonumber \\ [2mm]
& & + \frac{g_{\rm tc}^{2}}{M^{2}}
\left[ \overline{\psi _{L}}\, \left(\cos \theta \;\chi_{R}^{\prime }
-\sin \theta \;t_{R}^{\prime }\right)\right] 
\left[ \left(\cos\theta \; \overline{\chi_{R}^{\prime}}
-\sin\theta \; \overline{t_{R}^{\prime}} \right) \,\psi _{L}\right] 
\label{njl2}
\eear
where 
\be
\overline{M}=\sqrt{\mu_{\chi \chi }^{2}+\mu_{\chi t}^{2}} ~.
\label{barM}
\ee
We now proceed with the analysis by factoring the
interaction term in (\ref{njl2}) by introducing a static auxiliary
color-singlet field, $\varphi_0$ (which will become 
the {\it unrenormalized} composite Higgs doublet), to
obtain:
\be
{\cal {L}}_{0} =  {\cal {L}}_{\rm kinetic}-
\left[\overline{M}\,\overline{\chi_{R}^{\prime }}\,\chi_{L}
+ g_{\rm tc}\,\overline{\psi_{L}}\,(\cos \theta \;\chi_{R}^{\prime }-\sin\theta
\;t_{R}^{\prime })\,\varphi_0 + {\rm h.c.} \right]
-M^{2}\varphi_0^{\dagger }\varphi_0 ~.
\label{l1}
\end{equation}

We now derive the low energy effective Lagrangian by means of the
block-spin renormalization group. We view eq.~(\ref{l1}) as the effective
Lagrangian of the theory at a distance scale $\sim 1/M$. To derive the effective
Lagrangian at a larger distance scale, $\sim 1/\mu $, where $M>\mu $, we
integrate out the modes of momenta $M\geq |k| \geq \mu $. For $M>
\overline{M}>\mu $ the field $\chi $ decouples, and we obtain:
\begin{equation}
{\cal {L}}_{\overline{M}>\mu }= {\cal {L}}_{\rm kinetic} -
g_{\rm tc} \sin\theta \left( \overline{\psi _{L}}
t_{R}^{\prime} \,\varphi_0 +{\rm h.c.} \right) + 
Z_{\varphi}\left| D\varphi_0\right|^{2} -
\widetilde{M}_{\varphi_0}^{2}(\mu) \varphi_0^{\dagger }\varphi_0 -
\widetilde{\lambda} \left(\varphi_0^{\dagger }\varphi_0\right)^{2}
\label{renlag}
\end{equation}
In the limit $M>\overline{M}>\mu $, we obtain by integrating the fermion loops:
\bear
\widetilde{M}_{\varphi}^{2}(\mu) & = & 
M^{2}-\frac{g_{\rm tc}^{2}N_c}{8\pi ^{2}}\left[ M^{2}-
\cos^2\theta \; \overline{M}^{2}  
\ln \left( \frac{M^{2}}{\overline{M}^{2}}\right) \right] +
{\cal O}\left(\overline{M}^{2}, \mu^2 \right) ~,
\nonumber \\ [2mm]
Z_{\varphi} & = & \frac{g_{\rm tc}^{2}N_c}{16\pi ^{2}} \left[
\ln \left(\frac{M^{2}}{\overline{M}^{2}}\right) +
\ln \left( \frac{\overline{M}^{2}}{\mu ^{2}}\right) \sin^{2}\theta 
+ {\cal O}(1) \right] ~,
\nonumber \\ [2mm]
\widetilde{\lambda} & = & \frac{g_{\rm tc}^{4}N_c}{8\pi^{2}} \left[
\ln \left( \frac{M^{2}}{\overline{M}^{2}}\right) + 
\ln \left( \frac{\overline{M}^{2}}{\mu^{2}}\right) \sin^{4}\theta 
+ {\cal O}(1) \right] ~.
\label{renpar}
\eear
These relationships are true for $\overline{M}>\mu $ in the large $N_{c}$
approximation, and 
illustrate the decoupling of the $\chi $ field at the scale $\overline{M}.$
In the limit $\sin \theta \ll 1$ we see that
the induced couplings are those of the usual NJL model. However, in this
limit the Higgs doublet is predominantly a boundstate of $\overline{\chi_R}\psi_L$, 
and the corresponding loop, with loop-momentum ranging over $M> |k| >\overline{M}$,
controls most of the renormalization group evolution of the effective Lagrangian.

Consider, therefore, the limit $\sin^{2}\theta \ll 1$, hence 
$\cos^{2}\theta \approx 1$. In order for the composite Higgs doublet to develop a
VEV, the $SU(3)_{1}$ interaction must be supercritical. The criticality condition 
corresponds to demanding a negative $\widetilde{M}_{\varphi}^{2}(\mu) $ as 
$\mu \rightarrow 0$:
\begin{equation}
\frac{g_{\rm tc}^{2}N_c}{8\pi ^{2}}\geq 
\left[ 1- \frac{\mu_{\chi \chi}^2}{M^{2}} 
\ln \left( \frac{M^{2}}{\mu_{\chi \chi}^2} \right) \right]^{-1}
\end{equation}
This condition is equivalent to the NJL criticality condition for 
$\mu_{\chi\chi }^2/M^{2}\ll 1$.
Once we take $g$ to be supercritical, we are
free to tune the renormalized Higgs boson mass, $M_{\varphi}^{2}(\mu) =
\widetilde{M}_{\varphi}^{2}(\mu) /Z_{\varphi}$, to any desired value. 
This implies that we are free
to adjust the renormalized VEV of the Higgs doublet to the electroweak value,
$\langle \varphi^{0}\rangle = v/\sqrt{2} \approx 175$ GeV. The 
effective
Lagrangian at low energies, written in terms of the
renormalized field $\varphi$, takes the form:
\begin{equation}
{\cal {L}}_{\overline{M}>\mu }= {\cal {L}}_{\rm kinetic}
- g_t \sin\theta \left(\overline{\psi_{L}}
t_{R}^{\prime }\,\varphi+{\rm h.c.} \right) + \left| D\varphi\right|^{2}-
M_{\varphi}^{2}(\mu) \, \varphi^{\dagger}\varphi
-\lambda (\varphi^{\dagger}\varphi)^{2}
\end{equation}
where:
\begin{equation}
\varphi = \varphi_0\sqrt{Z_\varphi};\;\,;\quad
g_t = \frac{g_{\rm tc}}{\sqrt{Z_{\varphi}}} \;\,;\quad
M_{\varphi}^{2}(\mu) = \frac{\widetilde{M}_{\varphi}^{2}(\mu)}{Z_{\varphi}} \;\,;\quad
\lambda = \frac{\widetilde{\lambda }}{Z_{\varphi}^{2}}\,.
\end{equation}

The resulting top quark mass can be read off from the renormalized Lagrangian:
\begin{equation}
m_{t}= g_t \, \sin\theta \frac{v}{\sqrt{2}} ~,
\end{equation}
which corresponds to a Pagels-Stokar formula of the form:
\begin{equation}
v^{2}= \frac{N_c}{8\pi ^{2}}\frac{m_{t}^{2}}{\sin ^{2}\theta \;}
\ln \left( \frac{M^{2}}{\overline{M}^{2}}\right) + {\cal O}(\sin ^{2}\theta
)\,.
\label{ps1}
\end{equation}
The Pagels-Stokar formula differs from that obtained (in large $N_c$
approximation) for top quark condensation models by the large enhancement
factor $1/\sin ^{2}\theta$. This is a direct consequence of the seesaw
mechanism. 

We note that, in principle, using the freedom to adjust $\sin\theta$ we
could accommodate any fermion mass lighter than 600 GeV.  This freedom
may be useful in constructing more complete models involving all three
generations. The top quark is unique, however, in that it is very
difficult to accommodate such a heavy quark in any other way. We
therefore believe it is generic, in any model of this kind, that the top
quark receives the bulk of its mass through this seesaw mechanism.

To better understand the connection to the seesaw mechanism we can view the
dynamics of the top quark mass from the mixing with the $\chi $ field. The
mass matrix for the heavy charge 2/3 quarks takes the form:
\begin{equation}
\pmatrix{\overline{t_L}&\overline{\chi_L}\cr }\,
\pmatrix{0& m_{t\chi}\cr \mu_{\chi t}&\mu_{\chi\chi}\cr}
\pmatrix{t_R\cr \chi_R\cr} ~.
\label{f1}
\end{equation}
where $m_{t\chi }$ is
dynamically generated by the VEV of the composite Higgs, $\varphi$, thus
satisfying the Pagels-Stokar relationship:
\begin{equation}
v^{2}= \frac{N_c}{8\pi ^{2}}\, m_{t\chi }^{2}\ln \left( \frac{
M^{2}}{\mu_{\chi \chi }^{2}}\right) ~.
\label{ps2}
\end{equation}
If the logarithm is not very large, then we obtain the advertised value 
$m_{t\chi } \sim 600$ GeV.
Diagonalizing the fermionic mass matrix of (\ref{f1}) for 
$\mu_{\chi\chi} \gg m_{t\chi }$ leads to the physical top mass:
\begin{equation}
m_{t} \approx \frac{m_{t\chi }\mu_{\chi t}}{\mu_{\chi \chi }}= m_{t\chi }\tan
\theta ~,
\label{mttan}
\end{equation}
and substitution of (\ref{mttan}) into (\ref{ps2}) reproduces (\ref{ps1})
for small $\tan\theta\approx\sin\theta$.

The minimization of the Higgs potential gives the usual NJL result, that the Higgs
boson has a mass twice as large as the dynamically generated fermion mass, which is 
$m_{t\chi}$ in the present case. Thus, the schematic model includes only one composite 
Higgs boson, which is heavy, of order 1 TeV. In Section 3.4 we will show that
in a more general theory that includes the seesaw mechanism there are more composite 
scalars, and one of the neutral Higgs bosons may be as light as ${\cal O}(100$ GeV).

We note that the inclusion of the $b$-quark is straightforward, and the
schematic model affords a simple way to suppress the formation of a $b$-quark
mass comparable to the top quark mass. We include additional
fermionic fields of the form $\omega _{L}$, $\omega _{R}$, and $b_{R}$ with
the assignments:
\begin{equation}
b_{R}, \omega _{L} : \, ({\bf 1,3,1,} -2/3) \;\;\; , \;\;\; 
\omega _{R}   : \, ({\bf 3,1,1,} -2/3) ~.
\label{omega}
\end{equation}
These fermion gauge assignments cancel the anomalies noted above.
We further allow $\overline{\omega _{L}}\omega _{R}$ and  $\overline{\omega
_{L}}b_{R}$ mass terms, in direct analogy to the $\chi $ and $t$ mass terms:
\begin{equation}
{\cal {L}}_{0}\supset -(\mu_{\omega \omega }\overline{\omega _{L}}\omega
_{R}+\mu_{\omega b}\overline{\omega _{L}}b_{R}+{\rm h.c.})
\end{equation}
We can suppress the formation of the $\overline{\omega _{L}}b_{R}$
condensate altogether by choosing $\overline{M}_{\omega } =
\sqrt{\mu^2_{\omega\omega}+\mu^2_{\omega b}}\sim M$. In this limit we do
not produce a $b$-quark mass.  However, by allowing $\mu_{\omega \omega
  }\leq M$ and $\mu_{\omega b}/\mu_{\omega \omega }\ll 1$ we can form an
acceptable $b$-quark mass in the presence of a small 
$\overline{\omega_{L}}b_{R}$ condensate.  Yet another possibility arises within this
model, though it will not be a general feature of these schemes, i.e.,
to exploit instantons \cite{tc2}. If we suppress the formation of the
$\overline{\omega _{L}}b_{R}$ condensate by choosing
$\overline{M}_{\omega }\sim M,$ there will be a
$\overline{\omega _{L}}b_{R}$ condensate induced via the 't Hooft determinant
when the $t$ and $\chi$ are integrated out.  
We then estimate the scale of the induced $\overline{\omega _{L}}b_{R}$
mass term to be about $\sim 20$ GeV, and the $b$-quark mass then emerges as 
$\sim 20\mu_{\omega  b}/\mu_{\omega \omega }$ GeV. We will not further elaborate the 
$b$-quark mass in the present discussion, since its precise origin depends critically upon 
the structure of the complete theory including all light quarks and leptons.


\section{The Effective Potential Formalism}

\subsection{More General Interactions }
\setcounter{equation}{0}

Presently we extend the schematic model to include various additional
interactions, beyond the topcolor interaction of eq.~(\ref{njl0}). 
While we would ultimately like to replace
the $\mu_{\chi\chi}$ and $\mu_{\chi t}$ explicit mass terms 
exclusively with additional
strong dynamics, we find presently
that is not possible without the inclusion of additional
fields and additional dynamics.  The
seesaw mechanism, at least in the large-$N_c$ fermion loop approximations
seems to require these terms, and 
they also lift unwanted massless Nambu-Goldstone
bosons. In Section 4 we will sketch
out a more general high energy theory in
which these masses may arise dynamically, in analogy to
extended technicolor.
However, in the present case, these mass terms will be viewed as 
``small," 
in contrast to the schematic model in which they were large.

The NJL approximation illustrated in Section 2 
is probably a reasonable guide to
the physics of topcolor. One can frame the discussion
in terms of ``gap equations" and their solutions, 
as in \cite{dhseesaw}, but
it is useful and convenient to
have a more general and detailed description. In particular, the vacuum
structure of the topcolor theory is crucial to the success of the enterprise,
and it is important to study it with all the tools at our disposal. One of the
most useful tools is the effective potential \cite{eft}. This
has been used in \cite{bhl} to analyze simple topcolor models, and it was
employed in Section 2 in eqs.~(\ref{renlag}) and (\ref{renpar}) in lieu 
of the exclusive use of
gap equations as in \cite{dhseesaw}. In this Section
we extend its use in the present seesaw scheme
involving additional strong interactions . 

We thus consider a low energy effective theory, valid up to a scale $M >
O(10\ {\rm TeV})$, consisting of the standard model gauge group and fermions, 
and
a new vectorlike quark, $\chi$, which transforms under the
$SU(3)_C\times SU(2)_W \times U(1)_Y$ gauge group exactly as the right-handed
top, $t_R$. 

We assume that at the common scale $M$
the following four-fermion, NJL-like interactions, 
involving the top, bottom and
vectorlike quarks, occur:
\be
{\cal L}_{\rm int} = \frac{8\pi^2}{N_c M^2} \sum_{A,B = b, t, \chi} z_{AB}
\left(\overline{A}_L B_R\right)
\left(\overline{B}_R A_L\right) ~,
\label{ops1}
\ee
where $N_c = 3$ is the numbers of colors. The $z_{AB}$
($A,B = b, t, \chi$) are dynamical
coefficients determined by the couplings of
the high energy theory.
At the scale $M$ the electroweak symmetry is unbroken, implying 
$z_{bA} = z_{tA}$. Hence there are six independent $z_{AB}$ coefficients.
Our normalization is chosen so that the
interaction strength will be approximately critical (subcritical) 
in the $AB$ channel when
$z_{AB}>1 $ ($z_{AB}<1 $).
The interactions of eq.~(\ref{ops1}) should be viewed as Fierz-rearranged 
versions
of single massive gauge boson exchange interactions arising in a more general
high energy theory. For example,  
we imagine that the four-fermion operators (\ref{ops1}) arise from 
topcolor-like
\cite{topcolor} interactions, and therefore $ z_{AB}$ are functions of 
gauge couplings and charges. In the special case of the
schematic model of
Section 2, $z_{t\chi} = N_c g_{tc}^2/(8\pi^2)$ and all other $z_{AB}$ coefficients
are zero. In the model introduced in ref.~\cite{dhseesaw} all $z_{AB} \sim 1$, and 
their dependence upon the charges is given in the 
present paper in Appendix A [see eq.~(\ref{zab})].

In addition to the four-fermion operators (\ref{ops1}), small, explicit,
electroweak preserving mass terms are allowed in the Lagrangian:
\be
{\cal L}_{\rm mass} = - \mu_{\chi\chi} \overline{\chi}_L\chi_R
- \mu_{\chi t} \overline{\chi}_L t_R + {\rm h.c.}
\ee
The model presented in Section 4 is an example of high
energy physics that generates dynamically these four-fermion operators and masses.

\subsection{The Effective Potential}

The four-fermion interactions can be 
factorized, at the scale $M$,  by introducing  static auxiliary fields 
${\phi_0}_{AB} \equiv \bar{B}_R A_L$ ($A,B = b, t, \chi$),
which are described by the following effective Lagrangian:
\bear
{\cal L}_{\rm eff} & = & \sum_{A,B = b, t, \chi} \left[
\left(\overline{A}_L B_R {\phi_0}_{AB} + {\rm h.c.}\right) +
\frac{N_c M^2 }{8\pi^2 z_{AB}} {\phi_0}_{AB}^\dagger {\phi_0}_{AB}\right]
\nonumber \\ [2mm] & &
- \left( \mu_{\chi\chi} \overline{\chi}_L\chi_R
+ \mu_{\chi t} \overline{\chi}_L t_R + {\rm h.c.} \right) ~.
\eear
At the scale $M$, the ${\phi_0}_{AB}$ have
vanishing kinetic terms. 
At scales below  $M$ the ${\phi_0}_{AB}$ will acquire
kinetic terms through the effects of fermion loops
and  become propagating composite
scalars fields.  The loops also
generally induce running
mass terms and running quartic and Yukawa interactions.  The fields are
renormalized ${\phi_0}_{AB}\rightarrow \phi_{AB}$,
to give conventional kinetic term normalizations, 
and we thus find at a scale $\mu < M$, using, e.g., block spin
renormalization group in the large $N_c$ approximation, the
effective Lagrangian:
\be
{\cal L}_{\rm eff}^\mu = g_t \sum_{A,B = b, t, \chi}
\left(\overline{A}_L B_R \phi_{AB} + {\rm h.c.}\right) + 
\left(D_\nu \phi_{AB}^\dagger\right) \left(D^\nu \phi_{AB}\right) - V(\phi) ~.
\ee
Here we redefined the renormalized scalar fields by including
a shift to absorb the explicit mass fermionic terms:
\be
\phi_{AB} \equiv {\phi_0}_{AB} \sqrt{Z_\phi} - \frac{\mu_{AB}}{g_t} ~,
\ee
where $Z_\phi$ is the wave function renormalization,
and $\mu_{AB} = 0$, except for $\mu_{\chi\chi}$ and $\mu_{\chi t}$.
In the large $N_c$ limit, the one-loop effective potential is given by:
\be
V(\phi) = \frac{\lambda}{2} 
{\rm Tr}\left[\left( \phi^\dagger \phi\right)^2\right]
+ \sum_{A,B = b, t, \chi} \left[ M_{AB}^2 \phi_{AB}^\dagger \phi_{AB} 
+ C_{AB} \left( \phi_{AB} + \phi_{AB}^\dagger \right) \right] ~.
\label{effective}
\ee 
Note that the trace is just the sum over repeated indices of
$\phi_{AB}^\dagger \phi_{BC}\phi_{CD}^\dagger \phi_{DA}$.
The renormalized quartic and Yukawa coupling constants depend 
logarithmically on the physical cut-off,
\be
g_t = \sqrt{\frac{\lambda}{2}} =
\frac{4 \pi }{ \sqrt{N_c \ln \left(M^2/\mu^2\right)}} ~,
\ee
while the scalar squared-masses and tadpole coefficients 
depend quadratically on $M$:
\bear
M_{AB}^2 & = & \frac{2 M^2}{\ln \left(M^2/\mu^2\right)}
\left(\frac{1}{z_{AB}} - 1 \right) ~,
\nonumber \\ [3mm]
C_{AB} & = & \frac{\mu_{AB} M^2}{2\pi z_{AB}}
\sqrt{\frac{N_c }{\ln \left(M^2/\mu^2\right)}} > 0 ~.
\label{param}
\eear
Note that with our conventions the $C_{AB}$ are positive
and electroweak symmetry imposes 
$C_{Ab} = C_{tt} = C_{t\chi} = 0$.
This is just the usual effective potential derivation as in \cite{bhl}
applied to the present more general interaction.

In order to determine the vacuum properties of the theory 
we  minimize the
effective potential. Note that a global $U(1)_{b_R}$ symmetry
forbids tadpole terms for the $\phi_{Ab}$ scalars, independent of the
VEVs of the other scalars. We further assume that the $Ab$ channels are
subcritical, thus:
\be
 z_{Ab} < 1 \, , \, A = b, t, \chi ~,
\label{zbr}
\ee
so that $M_{Ab}^2 > 0$. As a result, the composite scalars having $b_R$ as
constituents do not acquire VEVs: $\langle \phi_{Ab} \rangle = 0$.
An $SU(2)_W$ transformation allows us to set
$\langle \phi_{b\chi} \rangle = 0$. We also take $z_{tt} < 1$, such that
$M_{tt}^2 > 0$, which implies that $\phi_{bt}$ and $ \phi_{tt}$ may acquire
VEVs only if they have tadpole terms induced by the VEVs of the other
scalars.
This implies that the VEVs of the $SU(2)_W$ doublet scalars, $\bar{t}_R \psi_L$
and $\bar{\chi}_R \psi_L$, are aligned, so that 
$\langle \phi_{bt} \rangle = 0$.
Finally, it is obvious that a nonzero VEV for $\phi_{t \chi}$ requires
$M_{t \chi}^2 < 0$, while the signs and sizes of $M_{\chi t }^2$
and $M_{\chi \chi}^2$ are not constrained so far.

Altogether only four out of the nine composite fields may have nonzero VEVs:
$\langle\phi_{AB}\rangle$ with $\, A,B = \chi, t$. 
At the minimum of the effective potential (\ref{effective}), the phases 
of the $\phi_{\chi\chi}$ and $\phi_{\chi t}$ are forced to be $-1$ 
by the tadpole terms (recall that the electroweak symmetry imposes 
$C_{tt} = C_{t\chi} = 0$). In addition, the relative phase 
between $\langle\phi_{t\chi}\rangle$ and $\langle\phi_{tt}\rangle$ has to 
be negative in order to minimize the quartic terms in the effective potential. 
Thus, there is only one arbitrary phase left, which can be 
fixed by choosing $\langle\phi_{tt}\rangle > 0$.
Let us denote the absolute values of the VEVs by $v_{AB}$, so that:
\be
v_{tt} = \langle \phi_{t t} \rangle  \; , \quad
 v_{\chi\chi} = -\langle \phi_{\chi\chi} \rangle \; , \quad
 v_{t \chi} = -\langle \phi_{t \chi} \rangle  \; , \quad
 v_{\chi t} = -\langle \phi_{\chi t} \rangle  \; . \quad
\ee
The values of $v_{AB}$ can be determined by minimizing the following function:
\bear
V(v_{AB}) & = & \frac{\lambda}{2}\left[
\left( v_{tt}^2 + v_{t\chi}^2 \right)^2 
+ \left( v_{\chi\chi}^2  + v_{\chi t}^2 \right)^2 
+ 2  \left( v_{tt}v_{\chi t} - v_{\chi\chi}v_{t\chi} \right)^2 \right]
\nonumber \\ [2mm]
& & + \sum_{A,B = t, \chi} M_{AB}^2 v_{AB}^2 - 2 C_{\chi\chi} v_{\chi\chi}
- 2 C_{\chi t} v_{\chi t}  ~.
\label{efpot}
\eear

We would like to find a vacuum that satisfies a general seesaw condition.
It is convenient to parametrize the VEVs as follows (up to phases and an
overall factor of $g_t$, this is just
the fermionic mass matrix):
\be
\left( \ba{rl} v_{t t} & v_{t \chi} \\
    v_{\chi t} & v_{\chi \chi} \ea \right)
= v_{\chi\chi}
   \left( \ba{rl} a b \epsilon & \epsilon \\
    b & 1 \ea \right) ~.
\label{seesawvac}
\ee
In terms of the dimensionless parameters $a$, $b$ and $\epsilon$ introduced 
here, the general seesaw condition reads:
\be 
0 < \epsilon < b < 1 \; , \quad 0 < a \ll \frac{1}{\epsilon} \; , \quad 
\epsilon \ll 1 ~.
\label{seesawcond}
\ee
The limit $a, b \ll 1$ corresponds to the seesaw condition used in
ref.~\cite{dhseesaw}. One can easily check that the stationarity conditions,
\be
\frac{\partial V}{ \partial v_{AB} } = 0 \; , \; \; \; A,B = t, \chi ~,
\label{stationarity}
\ee
have indeed a solution satisfying eqs.~(\ref{seesawcond}).
This solution is a stable minimum of the effective potential if and only if
all four eigenvalues of the second derivative of $V$ are positive at the
stationary point. Before computing the eigenvalues, we note that the equations
$\partial V/ \partial v_{\chi\chi} = 0$ and
$\partial V/ \partial v_{\chi t} = 0$ give $\epsilon$ and $b$ in terms
of $C_{\chi\chi}$, $C_{\chi t}$ and $M_{AB}^2$. As a consequence, the 
conditions $C_{\chi\chi}$, $C_{\chi t} > 0$, used in fixing the phases of the VEVs,
impose the following restrictions:
\be
M_{\chi t}^2, \, M_{\chi\chi}^2 > M_{t\chi}^2 \frac{1 + b^2}{1- \rho b^2} 
\left[1 + {\cal O}(\epsilon^2) \right] ~,
\label{restr}
\ee
where we defined:
\be
\rho \equiv \frac{- M^2_{t\chi}}{M^2_{tt}} > 0 ~.
\ee 
The other two stationarity conditions,
$\partial V/ \partial v_{tt} = 0$ and
$\partial V/ \partial v_{t \chi} = 0$, yield:
\bear
a &=& \rho \left[1 + {\cal O}(\epsilon^2) \right] ~,
\nonumber \\ [2mm]
v_{\chi\chi}^2 &=& \frac{- M_{t\chi}^2}{\lambda (1-\rho b^2) }
\left[1 - \epsilon^2 (1+ \rho^2 b^2)\frac{1- \rho(1 - 2\rho)b^2 }{(1-\rho b^2)^2}
 + {\cal O}(\epsilon^4) \right] ~.
\label{vcc}
\eear
These expressions allow us to write the second derivative of $V(v_{AB})$ as 
the following $4\times 4$ matrix:
\be
\partial^2 V(v_{AB}) = 2 \lambda v_{\chi\chi}^2
\left( \ba{cc} {\cal A}_1 + \epsilon^2 {\cal B}_1 + {\cal O}(\epsilon^4) 
& \epsilon {\cal B}_3 +  {\cal O}(\epsilon^3) \\
\epsilon {\cal B}_3^\top +  {\cal O}(\epsilon^3) 
& {\cal A}_2 + \epsilon^2 {\cal B}_2 + {\cal O}(\epsilon^4) 
\ea \right) \, ~,
\label{secderiv}
\ee 
where ${\cal A}_{1,2}$ and ${\cal B}_{1,2,3}$ are 2$\times 2$ real matrices
that depend only on $b$ and $M^2_{tt,t\chi,\chi t, \chi\chi }$.
Note that the rows and columns of 
$\partial^2 V(v_{AB})$ are arranged in eq.~(\ref{secderiv}) in the following order:
$v_{tt},v_{t\chi},v_{\chi t}, v_{\chi\chi}$.
Using the explicit form of ${\cal A}_{1,2}$, 
\bear
{\cal A}_1 & = & \left( \ba{cc} 1/\rho  &  - b \\
- b & \rho b^2 \ea \right) \; ~,
\nonumber \\ [3mm]
{\cal A}_2 & = & \left(\frac{1-\rho b^2}{-M^2_{t\chi}} \right)
{\rm diag}\left(M_{\chi\chi}^2, \, M^2_{\chi t} \right) + \,
\left( \ba{cc} 3 + b^2 &  2 b \\
2 b  &  1 + 3 b^2 \ea \right) ~,
\eear 
it is easy to compute to first order in $\epsilon^2$ the eigenvalues
of $\partial^2 V(v_{AB})$. Three of these are positive 
[eq.~(\ref{restr}) is important here], 
while the fourth eigenvalue cancels to leading order in $\epsilon^2$.
To ensure vacuum stability, the corrections of order $\epsilon^2$ to
$\partial^2 V$ must give a positive contribution to this eigenvalue.
We check this condition in Section 3.4, where we also show
that this eigenvalue corresponds to the mass of a light Higgs boson.

\subsection{Parameter Space}

The effective potential depends on six squared-masses
$M^2_{tA}, M^2_{\chi A}$ ($A = b,t,\chi$), two tadpole coefficients
$C_{\chi\chi}$, $C_{\chi t}$, and on $\ln (M/\mu)$.
We will choose the renormalization point $\mu$ to be the mass of the 
$\chi$ fermion. In doing so, we will neglect the running of the coefficients  
in the effective potential between the scale $m_\chi$ and the scale $m_t$.
In practice, this approximation is justified only if 
$M/m_\chi > m_\chi/m_t \sim 1/\epsilon$. We emphasize that this condition is 
not needed in a more developed computation of the renormalization group 
evolution.

We will proceed with deriving the constraints imposed on the parameters of the  
effective potential by the measured values of the $W$, $Z$ and $t$ masses.
The elements of the fermion mass matrix are proportional to the VEVs:
\be
m_{AB} = - g_t \langle\phi_{AB}\rangle ~.
\ee
It is straightforward to compute the top and $\chi$ quark masses
[see eq.~(\ref{mtmc})]:
\bear
m_t & = & m_{t\chi} \frac{b\left(1 + \rho \right)}{\sqrt{1 + b^2}} \,
\left[1 + {\cal O}\left( \epsilon^2 \right) \right] ~,
\nonumber \\ [2mm]
m_\chi & = & \frac{m_{t\chi}}{\epsilon} \sqrt{ 1 + b^2 }
\left[1 + \frac{\epsilon^2}{2} \left(\frac{1 - \rho b^2}{1 + b^2}\right)^2 +
 {\cal O}\left( \epsilon^4 \right) \right] ~.
\label{fmasses}
\eear
The electroweak symmetry is broken only by the VEVs of $\phi_{\psi \chi}$ 
and $\phi_{\psi t}$,
\be
\frac{v^2}{2} = v_{t\chi}^2 + v_{tt}^2 ~,
\ee
which implies:
\bear
m_{t\chi} & = & \frac{g_t v}{\sqrt{2(1 + \rho^2 b^2)}}
\left[1 + {\cal O}\left( \epsilon^2 \right) \right]
\nonumber \\ [2mm]
& \approx & 890 \; {\rm GeV} \;
\left[\left(1 + \rho^2 b^2\right) \ln \left(\frac{M}{m_\chi}\right) \right]^{\! -1/2} ~.
\eear
Using the expression for the top quark mass in eq.~(\ref{fmasses}), we find a 
constraint on $b$ and $\rho$,
\be
\frac{b^2 \left(1 + \rho \right)^2}{(1+ b^2)(1 + \rho^2 b^2)}  \approx
4\times 10^{-2} \ln \left(\frac{M}{m_\chi}\right) ~,
\label{bvalue}
\ee
which shows that $b^2 \lae {\cal O}(0.1)$ ($M$ is not larger by many orders 
of magnitude than $m_{\chi}$ unless
the coefficients of the four-fermion operators are excessively fine-tuned to be close 
to the critical value).

The $\chi_L^\prime$ mass eigenstate couples to $W$ and $Z$ so that there is a
potentially large custodial symmetry violation. However, in the decoupling
limit ($\epsilon/b \rightarrow 0$) this effect vanishes. To show this we
consider the one-loop contribution of $\chi$ to the $T$ parameter:
\be
T = \frac{3}{16 \pi^2 \alpha v^2}
\left[ s_L^4 m_\chi^2 + 2 s_L^2 (1- s_L^2)
\frac{ m_\chi^2 m_t^2}{ m_\chi^2 - m_t^2}
\ln \left(\frac{m_\chi^2}{m_t^2}\right) - s_L^2 (2 - s_L^2) m_t^2 \right] ~,
\label{tfromloop}
\ee
where $s_L$ is the sine of the left-handed mixing angle, defined in
eq.~(\ref{mix}):
\be
s_L = \epsilon \left[1 - b^2 \frac{(1 + \rho)(3- \rho) }{2 (1 + b^2)}
 \right]^{\! 1/2} + {\cal O}\left( \epsilon^3 \right) ~.
\ee
Because this mixing is small, the $\chi$ loop contribution to $T$ is
suppressed compared to the top loop contribution by a factor of
$\epsilon^2/b^2$:
\be
T = \frac{N_c m_t^2}{16 \pi^2 \alpha(M_Z^2) v^2} \frac{\epsilon^2}{b^2}
\left[ 1 - 4 b^2 \ln (\epsilon b)\right]
\left[ 1 + {\cal O}(b^2, \epsilon^2)\right] ~.
\ee
In practice, the current experimental constraints on $T$ are satisfied
if $b$ is larger than $\epsilon$ by a factor of 2 or so \cite{dhseesaw}.
Thus, the upper bound on $\epsilon$ is about 0.1, confirming 
that the expansion in $\epsilon^2$ is a good approximation.

To summarize, for $M/m_\chi \sim 10$ the elements of the fermion mass
matrix eq.~(\ref{seesawvac}) have sizes:
\be
\pmatrix{
m_{tt} \lae {\cal O}(100\ {\rm GeV}) & m_{t\chi} \sim {\cal O}(600\ {\rm GeV}) \cr
m_{\chi t}\gae {\cal O}(1\ {\rm TeV}) & m_{\chi\chi} \gae {\cal O}(5\ {\rm TeV})
}~.
\ee \\ [-2mm]
The effective potential analysis given is valid only for $M \gg
m_{\chi\chi}$. Smaller values of $M$ (with less fine-tuning) may be
allowed, though we cannot demonstrate that fact.  The relations between
$\epsilon,b$ and $C_{\chi\chi}, C_{\chi t}$ allow us to estimate the
$\mu_{\chi\chi}$ and $\mu_{\chi t}$ mass coefficients from the
Lagrangian:
\be 
\mu_{\chi A} = m_{\chi A} \, \frac{M_{\chi A}^2 -
  M_{t\chi}^2}{2 z_{\chi A} M^2}\, \ln \left(\frac{M}{m_\chi}\right) \left[1 +
{\cal O}\left( b^2, \epsilon^2 \right) \right] ~.
\label{mus}
\ee
Generically we expect $ |M_{AB}| \sim {\cal O}(m_\chi) < \epsilon M$, so that 
$\mu_{\chi A}/m_{\chi A} < {\cal O}(\epsilon^2)$. By contrast, in Section 2
the schematic model does not lead to a $\phi_{\chi t}$ or $\phi_{\chi \chi}$
bound state, and eq.~(\ref{mus}) is replaced by $\mu_{\chi A} = m_{\chi A}$.

\subsection{The Composite Scalar Spectrum}

Next we compute the composite scalar spectrum. The 3$\times$3 matrix
$\phi$ contains a total of 18 real scalar degrees of freedom,
corresponding potentially to a Higgs sector which includes three
weak-doublets, $\phi_{\psi A} \equiv \overline A_R \psi_L$ with $A = b,
t, \chi$ and $\psi_L = (t,b)_L$, and three weak-singlets, $\phi_{\chi A}
\equiv \overline A_R \chi_L$. 

An unbroken global $U(1)_{b_R}$ symmetry ensures that the
$\phi_{\psi b}$ and $\phi_{\chi b}$ scalars do not mix with 
$\phi_{At}$ or $\phi_{A\chi}$.
Therefore, the neutral complex scalar $\phi_{bb}$ has a mass $M_{tb}$
given by eq.~(\ref{param}), and the complex  scalars $\phi_{tb}$ and $\phi_{\chi b}$
with electric charge +1 have a mass matrix: 
\be
{\rm diag}\left(M_{tb}^2, \, M^2_{\chi b} \right) + \,
\lambda v_{\chi\chi}^2 \left(\ba{cc} 
\epsilon^2 (1 + a^2 b^2) & \epsilon (1 - a b^2) \\ [2mm]
\epsilon (1 - a b^2) & 1 + b^2 \ea \right) ~.
\ee
In Section 3.2 we imposed $M_{tb}^2, M_{\chi b}^2 > 0$, which implies 
that the mixing between $\phi_{tb}$ and $\phi_{\chi b}$ is suppressed by $\epsilon$.
We will denote the mass eigenstates by $H_{tb}^\pm$ and $H_{\chi b}^\pm$.
The magnitudes of the masses that appear in the 
effective potential, $|M_{AB}|$, are expected to be roughly 
of the same order in the absence of fine-tuning. 
Using the relation:
\be
\lambda v_{\chi\chi}^2 = \frac{2}{\epsilon^2} m_{t\chi}^2
\ee
we can estimate $|M_{t\chi}|$ from eq.~(\ref{vcc}): 
\be
- M_{t\chi}^2 = \frac{2}{\epsilon^2} m_{t\chi}^2 \left(1-\rho b^2\right) 
\left[1 + {\cal O}(\epsilon^2) \right]
\ee
Given that $\rho b^2 \lae {\cal O}(0.1)$, as can be seen from 
eq.~(\ref{bvalue}), it follows that $|M_{t\chi}| \! \gae \! {\cal O}(5\ {\rm TeV})$.
If $M_{tb}$ is indeed of the same order as $|M_{t\chi}|$, then the two charged
scalars have masses of a few TeV or larger. On the other hand, if $z_{tb}$ and
$z_{\chi b}$ are tuned sufficiently close to one so that 
$M_{tb}, \, M_{\chi b} \ll b |M_{t\chi}|$,
then the mass eigenstate which is predominantly $\phi_{tb}$ has a mass-squared:
\be
M_{H_{tb}^\pm}^2 \approx 2 m_t^2 + M_{tb}^2 ~.
\ee
This sets a lower bound on the charged Higgs mass of about 250 GeV.

The other two complex scalars with electric charges $+ 1$, 
$\phi_{bt}$ and $\phi_{b\chi}$ have the following mass matrix:
\be
{\rm diag}\left(M_{tt}^2, \, M^2_{t\chi} \right) +\,
\lambda v_{\chi\chi}^2 \left(\ba{cc} b^2 (1 + a^2 \epsilon^2) &
b (1 - a \epsilon^2) \\ [2mm] b (1 - a \epsilon^2) &
1 + \epsilon^2 \ea \right) ~.
\ee\\ 
One of the eigenvalues vanishes, corresponding to the charged Nambu-Goldstone bosons
that become the longitudinal $W$. The other eigenvalue is the mass-squared of
a charged Higgs boson, $H^\pm$, and can be computed without expanding in powers of 
$\epsilon$ by using the stationarity conditions:
\be
M^2_{H^\pm} = \frac{2 m_{t\chi}^2}{a \epsilon^2} (1 + a^2 b^2)(1-a \epsilon^2) ~.
\ee
This mass is also large, most likely above a TeV.

There are four CP-even neutral scalars, ${\rm Re}\, \phi_{tt}$, 
${\rm Re}\, \phi_{t\chi}$, ${\rm Re}\, \phi_{\chi\chi}$ and 
${\rm Re}\, \phi_{\chi t}$. Their mass matrix is given by:
\be
\frac{1}{2} {\rm diag}(1, -1, -1, -1) \, \partial^2 V(v_{AB}) \, 
{\rm diag}(1, -1, -1, -1) ~,
\ee
with $\partial^2 V$ indicated in eq.~(\ref{secderiv}).
It is possible to compute the eigenvalues of this mass matrix as an 
expansion in $\epsilon^2$. 
There are two mass eigenstates which, to leading order in $\epsilon$,
are linear combinations of only ${\rm Re}\, \phi_{tt}$
and ${\rm Re}\, \phi_{t\chi}$. Since the electroweak symmetry is broken only by
the VEVs of $\phi_{tt}$ and $\phi_{t\chi}$, it is appropriate to label
these mass eigenstates by $h^0$ and $H^0$, as in a two Higgs doublet model:
\bear
h^0 & = & \sqrt{2}\left(1 + \rho b^2 \right)^{\! -1/2} 
\left({\rm Re}\, \phi_{t\chi} + b \sqrt{\rho} \, {\rm Re}\, \phi_{tt} \right) 
+ {\cal O}(\epsilon)
\nonumber \\ [2mm]
H^0 & = & \sqrt{2}\left(1 + \rho b^2 \right)^{\! -1/2} 
\left(- b \sqrt{\rho} {\, \rm Re}\, \phi_{t\chi} + {\rm Re}\, \phi_{tt} \right)
 + {\cal O}(\epsilon) ~.
\label{higgstate}
\eear
The electroweak symmetry is unbroken in the $\epsilon \rightarrow 0$ limit,
so that the heavy neutral Higgs boson is degenerate with $H^\pm$:
\be
M^2_{H^0} = \frac{2 m_{t\chi}^2 }{\rho \epsilon^2} \left(1 + \rho^2 b^2 \right)
\left[1 + {\cal O}(\epsilon^2)\right] 
= M^2_{H^\pm} \left[1 + {\cal O}(\epsilon^2)\right] ~.
\label{largemass}
\ee
It is easier to compute the mass of the lightest neutral Higgs 
boson, $M_{h^0}$, 
as a power series in $b^2$, which is a reasonably small parameter due to 
the constraint (\ref{bvalue}). The result is: 
\be
M^2_{h^0} = 4 m_{t\chi}^2 \,
\frac{M^2_{\chi\chi} - M^2_{t\chi}}{M^2_{\chi\chi} - 3 M^2_{t\chi}}
\left[1 + {\cal O}(b^2,\epsilon^2)\right] ~.
\ee
For $M^2_{\chi\chi} \sim - M^2_{t\chi}$, the $h^0$ is heavy, with a mass
of order $\sqrt{2} m_{t\chi} \sim 800$ GeV. In the schematic model presented in Section 2,
the $\phi_{\chi\chi}$ boundstate does not form, so that 
$M^2_{\chi\chi} \rightarrow \infty $ and we recover the NJL result $M_{h^0} = 2  m_{t\chi}$.
On the other hand, if $M^2_{\chi\chi} < 0$, the $h^0$ can be significantly lighter.
A composite neutral Higgs boson with mass of order 100 GeV would require 
a cancellation between  $M^2_{\chi\chi}$ and $M^2_{t\chi}$ at the level of 15\%.
Such a cancellation does not necessarily require fine-tuning: for instance, if 
$\psi_L$ and $\chi_L$ have the same charges under the broken gauge groups that
induce the four-fermion operators, then $z_{\chi\chi} = z_{t\chi}$ implying
$M_{\chi\chi} = M_{t\chi}$.
This shows that the existence of a light composite neutral Higgs boson, 
with a mass of order 100 GeV is a possibility. 

To leading order in $\epsilon$, the other two CP-even neutral mass eigenstates are 
linear combinations of ${\rm Re}\,\phi_{\chi\chi}$ and ${\rm Re}\,\phi_{\chi t}$, 
with a mixing of order $b$. Their squared-masses are given by:
\bear
M^2_{H^0_{\chi t}} & = & \frac{2  }{\epsilon^2}  m_{t\chi}^2
\left( 1 + \frac{M^2_{\chi t}}{-M^2_{t \chi}} \right)
\left[1 + {\cal O}(b^2, \epsilon^2)\right] ~,
\nonumber \\ [2mm]
M^2_{H^0_{\chi\chi}} & = & \frac{2 }{\epsilon^2}   m_{t\chi}^2
\left( 3 + \frac{M^2_{\chi\chi}}{-M^2_{t \chi}} \right)
\left[1 + {\cal O}(b^2, \epsilon^2)\right]  ~.
\eear 
The $H^0_{\chi\chi}$ is heavy, with a mass of at least ${\cal O}(5$  TeV), 
while $H^0_{\chi t}$ can be light, with a mass of order $m_{t\chi}$ or lower,
 if $M^2_{\chi t}$ and $M^2_{\chi\chi}$ are
close to their lower bound (\ref{restr}).
It is clear now that for typical values of the parameters in the effective potential 
all four CP-even neutral mass eigenstates have positive squared-masses,
which proves that the minimization of the effective potential performed in
Section 3.2 is correct. On the other hand, if the restriction (\ref{restr})
on $M^2_{\chi t}$ and $M^2_{\chi\chi}$ is saturated at order $\epsilon^2$,
then the masses of $H^0_{\chi t}$ or $h^0$ might vanish, signaling a second order
phase transition to an unacceptable vacuum.

The remaining four states are the CP-odd neutral scalars:
${\rm Im}\,\phi_{tt}$, ${\rm Im}\,\phi_{t\chi}$,
${\rm Im}\,\phi_{\chi\chi}$ and ${\rm Im}\,\phi_{\chi t}$.
In the $\epsilon \rightarrow 0$ limit the masses of the $\phi_{\psi t}$ and $\phi_{\psi\chi}$
doublets are $SU(2)_W$ invariant, so that the linear combination of 
${\rm Im}\,\phi_{tt}$ and ${\rm Im}\,\phi_{t\chi}$ analogous to $H^0$ in 
eq.~(\ref{higgstate}), labeled $A^0$, has a large mass given by eq.~(\ref{largemass}).
The other linear combination is the Nambu-Goldstone boson that becomes the longitudinal
$Z$. At order $\epsilon$, the longitudinal $Z$ includes a mixture of 
${\rm Im}\,\phi_{\chi\chi}$ and ${\rm Im}\,\phi_{\chi t}$.
The other two CP-odd mass eigenstates, $A^0_{\chi\chi}$ and $A^0_{\chi t}$ 
are predominantly ${\rm Im}\,\phi_{\chi\chi}$ and ${\rm Im}\,\phi_{\chi t}$,
respectively, and have large masses:
\be
M^2_{A^0_{\chi\chi}, \, A^0_{\chi t}} = \frac{2}{\epsilon^2}   m_{t\chi}^2 
\left[1 + b^2 + \frac{M^2_{\chi\chi, \,\chi t}}{-M^2_{t \chi}} 
\left(1 - \rho b^2\right)
+ {\cal O}(\epsilon^2)\right] ~.
\ee
These two neutral mass eigenstates are the pseudo Nambu-Goldstone bosons 
discussed in ref.~\cite{dhseesaw}, and are light provided 
$M^2_{\chi t}$ and $M^2_{\chi\chi}$ are close to  their bound (\ref{restr}).

The composite scalar spectrum has several features which warrant further comments.
The typical scale for the masses of the physical states corresponding to the two 
weak-doublets and two weak singlets which acquire VEVs is given by 
$m_{\chi\chi} = m_{t\chi}/\epsilon$. By contrast,
the $h^0$ has a mass proportional to $m_{t\chi}$, so that is a light state in the 
limit $\epsilon \rightarrow 0$. The reason for this result is the fact that 
the unitarity of the $WW$ scattering cross-section requires a state 
of the order of the electroweak scale, and 
the electroweak symmetry breaking VEVs, $v_{t\chi}$ and  $v_{tt}$,
are suppressed by a factor of $\epsilon$ compared with the other VEVs.
Therefore, the upper bound on the standard model 
Higgs boson mass, of order 1 TeV \cite{lqt}, is automatically 
enforced within  our 
composite Higgs sector.

The further suppression which allows $M_{h^0} \ll 1$ TeV  when 
$M^2_{\chi\chi} \approx M^2_{t \chi}$ is of a different nature. To see this, 
one should recall that, to leading order in $b^2$,
$C_{\chi\chi} \rightarrow 0$ (and also $\mu_{\chi\chi}\rightarrow 0$)
when $M^2_{\chi\chi} \rightarrow M^2_{t \chi}$ [see eq.~(\ref{restr})].
In this case, decreasing $M^2_{\chi\chi}$ triggers a second order
phase transition from the viable vacuum discussed thus far, to a new minimum
of the effective potential where only the weak-singlet fields
$\phi_{\chi t}$ and $\phi_{\chi\chi}$ 
have nonzero VEVs. The $h^0$ mass is therefore controlled by the proximity of 
$M^2_{\chi\chi}$ to the critical point. Note that we computed the lightest 
Higgs boson mass only to leading order in $b^2$, so that it is not clear
whether the phase transition is truly second order or 
weakly first order (in which case
there is a theoretical lower bound on $M_{h^0}$, but that may be below
the experimental bound of $\sim$ 100 GeV).  

It is also easy to understand why both $h^0$ and $A^0_{\chi\chi}$ 
have squared-masses
proportional to $M^2_{\chi\chi} - M^2_{t \chi}$ 
[or equivalently, to $\mu_{\chi\chi}$ as follows from eq.~(\ref{mus})]
in the limit of small $b^2$.
When $\mu_{\chi\chi}\rightarrow 0$, $A^0_{\chi\chi}$ 
becomes the Nambu-Goldstone
boson associated with a global $U(1)_\chi$ symmetry 
broken spontaneously by the $v_{\chi\chi}$ VEV, while the $h^0$ is the order parameter
of a second order phase transition.

To summarize, the composite scalar spectrum consists of the longitudinal
$W$ and $Z$ and the following states:

\begin{itemize}

\item $h^0$: \ a neutral Higgs boson of mass $m_{t\chi}\sim 600$ GeV times a factor 
of order one (or smaller if $M^2_{\chi\chi} \approx M^2_{t \chi}$);

\item $H^0,H^\pm,A^0$: \ the heavy states of a two Higgs-doublet sector,
roughly degenerate with a mass $ (m_{t\chi}/\epsilon)\sqrt{2/\rho}$; 

\item $H^0_{\chi t}, A^0_{\chi t}$: \ one CP-even and one CP-odd state, 
which are light only if $M^2_{\chi t} \approx M^2_{t \chi}$; 

\item $A^0_{\chi\chi}$: \ a neutral CP-odd state which is light only if 
$M^2_{\chi\chi} \approx M^2_{t \chi}$.

\item $\phi_{bb}$: \ a neutral complex scalar, with a mass $M_{tb}$ 
(which is an arbitrary parameter);

\item $H^\pm_{tb}$: \ a charged scalar which can be as light as 250 GeV 
if $M_{tb}$ and $M_{\chi b}$ are sufficiently small;

\item $H^0_{\chi\chi}, H^\pm_{\chi b}$: a CP-even neutral state and a charged
scalar, with large masses,$\gae \! m_{t\chi}/\epsilon$.
\end{itemize}
Finally we note that, for a generic choice of parameters, one or more of
these scalars may have a mass of order the cutoff, $M$. If so, these
particles are not part of the low-energy effective theory.

\section{Higher Energy Physics}
\setcounter{equation}{0}

We have shown in the previous Section that the top quark seesaw
mechanism leads to a low-energy effective theory involving bound states
of the $\chi$, $t$ and $b$ quarks. There are several questions that
remain: What breaks the topcolor gauge group? What
interactions distinguish $\chi$, $t$ and $b$? How is electroweak
symmetry breaking communicated to the other quarks and leptons?  In this
Section we describe a class of models of electroweak flavor symmetry breaking
incorporating a top quark seesaw which illustrates some of the issues
involved in constructing more complete models.

In the prototype model, topcolor symmetry breaking will be dynamically generated
while flavor symmetry breaking will be assumed to arise from unspecified
``extended topcolor'' interactions (analogous to extended technicolor
interactions \cite{etc}) at higher energies.  The model is most easily
displayed in ``moose notation'' \cite{moose}, in which lines stand for
fermion fields and circles for $SU(n)$ gauge groups. An arrow emerging from a
circle with an $n$ in it represents a left-handed fermion transforming like
$n$ or a right-handed fermion transforming like $\bar{n}$, while an arrow
going in indicates a right-handed fermion transforming like $n$ or a
left-handed fermion transforming like $\bar{n}$.

Using this notation, the prototype model is shown in Fig.~1.
\begin{figure}[htbp]
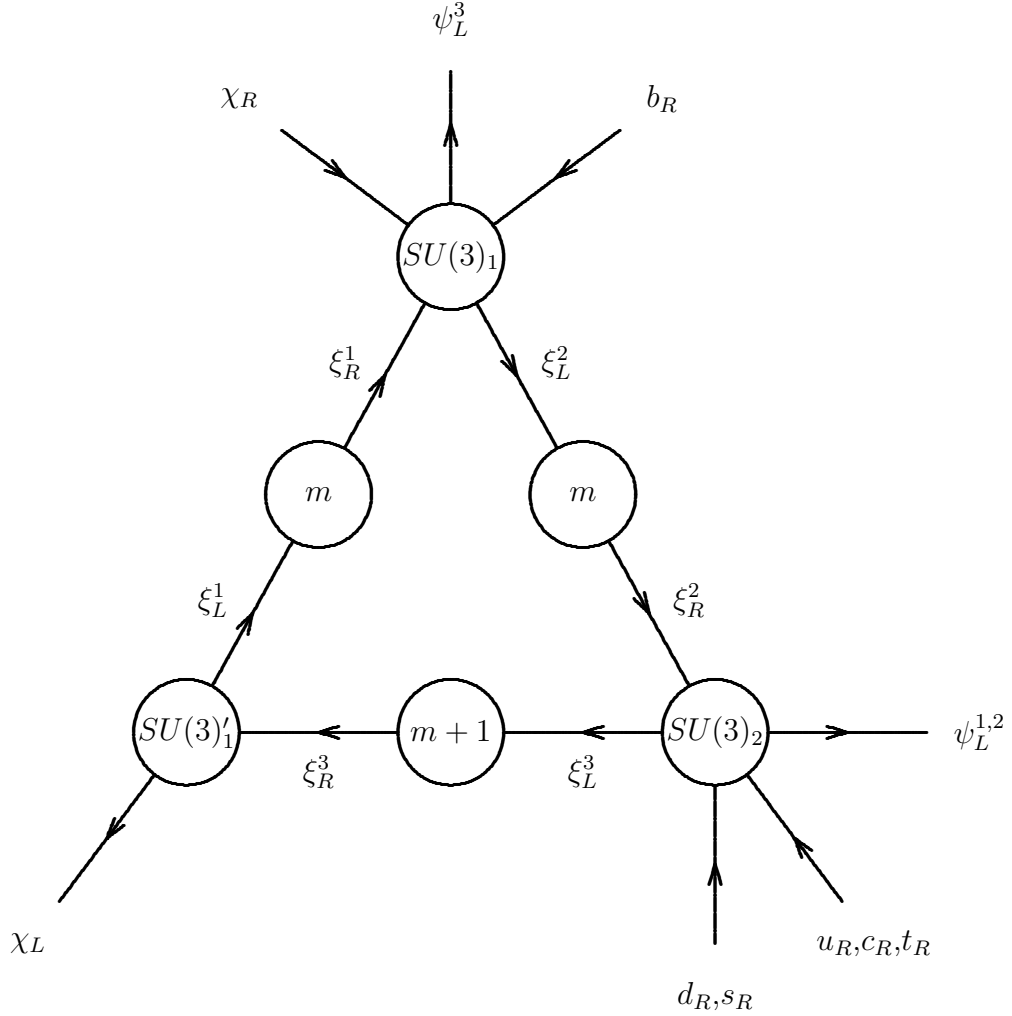

\centering
\hspace*{1cm}\parbox{5.5in}{
\beginpicture
\setcoordinatesystem units <\tdim,\tdim>
\stpltsmbl
\circulararc 360 degrees from 100 20 center at 100 0
\circulararc 360 degrees from 0 20 center at 0 0
\circulararc 360 degrees from -100 20 center at -100 0
\circulararc 360 degrees from 20 180 center at 0 180
\circulararc 360 degrees from 70 90 center at 50 90
\circulararc 360 degrees from -70 90 center at -50 90
\put {$SU(3)_1$} at 0 180
\put {$SU(3)^\prime_1$} at -100 0
\put {$SU(3)_2$} at 100 0
\moose{-20 0}{-50 0}{-80 0}
\moose{80 0}{50 0}{20 0}
\moose{-90 18}{-75 45}{-60 72}
\moose{-40 108}{-25 135}{-10 162}
\moose{10 162}{25 135}{40 108}
\moose{60 72}{75 45}{90 18}
\put {$m$} at 50 90
\put {$m$} at -50 90
\put {$m+1$} at 0 0
\moose{-112 -16}{-130 -40}{-148 -64}
\put {$\chi_L$} at -160 -80
\moose{100 -80}{100 -50}{100 -20}
\put {$d_R$,$s_R$} at 100 -100
\moose{148 -64}{130 -40}{112 -16}
\put {$u_R$,$c_R$,$t_R$} at 160 -80
\moose{120 0}{150 0}{180 0}
\put {$\psi^{1,2}_L$} at 200 0
\moose{-64 228}{-40 210}{-16 192}
\put {$\chi_R$} at -80 240
\moose{64 228}{40 210}{16 192}
\put {$b_R$} at 80 240
\moose{0 200}{0 230}{0 250}
\put {$\psi^3_L$} at 0 270
\put {$\xi^1_R$} at -40 140   
\put {$\xi^1_L$} at -90 50   
\put {$\xi^2_L$} at 40 140   
\put {$\xi^2_R$} at 90 50
\put {$\xi^3_L$} at 50 -15
\put {$\xi^3_R$} at -50 -15
\endpicture }
\caption{ The ``moose'' model of dynamical topcolor symmetry breaking.}
\label{model}
\end{figure}
The $\chi_{L,R}$ fields and right-handed quark fields are shown
explicitly, while $\psi^{1,2,3}_L$ denote the three generations
left-handed weak-doublet quark fields. We will assume here that the two
$SU(m)$ interactions and the $SU(m+1)$ interactions become strong and
produce $\bar{\xi}\xi$ condensates. The (relatively) strong $SU(3)_1$
interactions and the weaker $SU(3)_2$ gauge group are as in the schematic
model of Section 2: $SU(3)_1 \times SU(3)^\prime_1 \times SU(3)_2 \to
SU(3)_{QCD}$ due to the formation of $\bar{\xi^1_R}\xi^1_L$ and
$\bar{\xi^2_R} \xi^2_L$ condensates driven by a strong $SU(m)$ gauge
interactions, and a $\bar{\xi^3_R}\xi^3_L$ condensate driven by a strong
$SU(m+1)$ gauge interaction. 

The scale of $SU(3)_1 \times SU(3)^\prime_1$ breaking (set by the
$\bar{\xi}_L \xi_R$ condensates, {\it i.e.} the scales at which the two
$SU(m)$ interactions and the $SU(m+1)$ interaction become strong) is
assumed to be close to the scale at which the $SU(3)_1$ interactions
would break the chiral symmetries associated with the $\chi_R$ and
$\psi^3_L$ fields.  If that chiral phase transition is second-order,
this breaking gives rise to a $\bar{\chi_R}\psi^3_L$ composite Higgs
field.

The $\mu_{\chi \chi}$ and $\mu_{\chi t}$ ``mass'' terms cannot be
present at tree-level since the corresponding mass operators are not
gauge-invariant.  Instead, they must arise from higher-dimensional
operators coming from higher-energy interactions.  A
$\bar{\chi_L}\chi_R$ mass term can arise from an operator of the form
\begin{equation}
\bar{\chi_L}\gamma^\mu\xi^1_L\;\bar{\xi^1_R}\gamma_\mu\chi_R~,
\end{equation}
giving
\begin{equation}
\mu_{\chi\chi} \propto \langle \bar{\xi^1}\xi^1 \rangle~,
\end{equation}
while a $\bar{\chi_L}t_R$ mass term can arise from a four-fermion operator
of the form
\begin{equation}
\bar{\chi_L}\xi^3_R\;\bar{\xi^3_L}t_R~,
\label{chit}
\end{equation}
giving
\begin{equation}
\mu_{\chi t} \propto \langle \bar{\xi^3}\xi^3 \rangle~.
\end{equation}
As these ``masses'' are proportional to different condensates, their
sizes can naturally be different even if the sizes and strengths
of the corresponding higher-energy interactions are similar.
Furthermore, operators of the form shown in eq.~(\ref{chit}) can involve
all three generations of charge 2/3's quarks and is a potential
source of mixing between the third generation and the first two.

A crucial feature of the seesaw mechanism is that the $\overline \psi_L t_R$ 
mass term must be suppressed. This happens naturally in the model shown in 
Fig.~1: no gauge-invariant four-fermion operator exists which
could give rise to such a term. The largest contributions come from
six-fermion operators and are naturally small.

The masses and mixings of the first two
generations can easily arise from higher-energy interactions as well,
since both the left-handed and right-handed quarks transform under the
$SU(3)_2$ interactions. For example, a charm-quark mass can arise from
an operator of the form
\begin{equation}
\bar{\psi^3_L}\chi_R\,\bar{c_R}\psi^2_L~.
\end{equation}

The $b$ mass, however, is quite different here than in the schematic
model.  Because of the presence of the $\xi^1_R$ and $\xi^2_L$ fields
which also transform under $SU(3)_1$, instanton effects yield
high-dimension multifermion operators which are too small to account for
the bottom-quark mass.  We believe this will remain true
in any model of dynamical topcolor symmetry breaking.  Thus we have
assumed, counter-intuitively, that the $b_R$ shares topcolor
interactions with $\psi^3_L$ and $\chi_R$ so that we can allow for the
operator
\begin{equation}
\epsilon^{\alpha\beta}\,\bar{\psi^3_{L\alpha}}\chi_R\,\bar{\psi^3_{L\beta}}b_R
\end{equation}
(the $\epsilon^{\alpha\beta}$ acts on the $SU(2)_W$ indices to make it a
singlet). 
In addition to a $b$-quark mass, this operator induces a tadpole term for
$\phi_{bb}$ in the effective potential. However, the shift in the vacuum 
is small if $M_{tb}$ is large, and the analysis in Section 3 remains essentially 
unaltered.

Having given the $\chi_R$ and the $b_R$ the same strong gauge
interaction quantum numbers, we must introduce additional interactions
to ``tilt'' the vacuum and prevent the formation of a potentially large
$\bar{b_R} \psi^3_L$ condensate and a large bottom-quark mass. In the
spirit of extended technicolor, we will assume that the effective 
Lagrangian includes  operators like
\begin{equation}
\frac{\eta_\chi}{M^2} \bar{\psi^3_L}\chi_R\,\bar{\chi_R}\psi^3_L
+ \frac{\eta_b}{M^2} \bar{\psi^3_L}b_R\,\bar{b_R}\psi^3_L ~,
\label{tiltingi}
\end{equation}
with $\eta_\chi > \eta_b$. Such a pattern of interactions can tilt
the vacuum, as required. The presence of the operators in eq.~(\ref{tiltingi}) 
give rise to contributions to the $T$ parameter \cite{cdt}, beyond those
in eq.~(\ref{tfromloop}) coming from fermion loops. However, due to the 
large scale $M \sim {\cal O}(50$ TeV), these contributions are negligible
\cite{dhseesaw}. The same argument applies in the case of other electroweak 
observables \cite{ct} or FCNC effects \cite{fcnc}.

While we have yet to complete a full phenomenological analysis of this
model, we regard it as an existence proof that it is possible to
construct a model incorporating a top quark seesaw mechanism in which
topcolor symmetry breaking is dynamical and which allows for
intergenerational mixing. This model also raises additional questions:
What gives rise to the necessary higher-energy interactions?  Is there a
natural explanation for the near equality of the chiral symmetry
breaking scales of the $SU(m)$ and $SU(m+1)$ interactions? Why are these
chiral symmetry breaking scales close to the scale of $SU(3)_1$ chiral
symmetry breaking?

Finally, we note that a variant of this model could be constructed by
replacing the $b_R$ fermions transforming under $SU(3)_1$ by the
$\omega_R$ fermions of eq.~(\ref{omega}), adding the $b_R$ to the fields
transforming under $SU(3)_2$, and adding the $\omega_L$ to the fields
transforming under $SU(3)^\prime_1$. Anomaly cancellation will then
also require that $SU(m+1)$ is replaced by $SU(m+2)$. Such a variant allows for
additional sources of mixing between the third generation and the first
two. 

\section{Conclusions}
\setcounter{equation}{0}

In the dynamical top quark seesaw mechanism EWSB occurs via the
condensation of the left-handed top quark with a new, right-handed
weak-singlet quark. The fermionic mass scale of this weak $I=1/2$
condensate is large, of order 0.6 TeV, and it corresponds to the
formation of a dynamical boundstate Higgs scalar with a VEV 
$v/\sqrt{2} \approx 175$ GeV.  However, the new
$\chi$-quarks also condense amongst themselves at still larger scales,
and have condensates with the right-handed top quark as well. Upon
diagonalization of the fermionic mass matrix, the physical top quark mass
is suppressed compared to the 0.6 TeV matrix element by a seesaw mechanism.
As a result, this class of models allows for a dynamical origin of EWSB and can
accommodate a heavy top quark without introducing extra fermions carrying
weak-isospin quantum numbers.

In this paper we presented a schematic model with a minimal version of
the seesaw which illustrates the essential features of the dynamics. We
also presented a calculation of the effective potential in a generic 
low energy theory that incorporates the dynamical top quark seesaw mechanism.
This effective potential allows one to
understand the range of parameters required for the seesaw mechanism to
be successful. Furthermore, we have computed the spectrum of composite 
scalars, which includes a potentially light Higgs boson.
Finally, we presented class of models of electroweak
symmetry breaking which incorporate the top quark seesaw mechanism
and in which topcolor symmetry breaking is dynamically generated.

Many issues remain to be explored. Among these are: What is the most
elegant method to incorporate the first two generations of quarks and
intergenerational mixing, as well as leptons? Is there a natural
mechanism for topcolor to break close to its chiral symmetry breaking
scale? Are there generic experimental signatures of the top quark
seesaw?  We believe that the top quark seesaw opens up a wide range of
directions in model building which may allow these questions to be
answered.

\subsection*{Acknowledgments}

We are grateful to Bill Bardeen for stimulating discussions
regarding the schematic version of the seesaw and the existence of a 
light composite Higgs boson. 
We would like to thank Gustavo Burdman, Nick Evans, Deog Ki Hong,
Paul Mackenzie and Elizabeth Simmons for useful conversations.  
We also thank the Aspen Center 
for Physics for its hospitality during early stages of this work.

\section*{Appendix A: The $U(1)$ Tilting Model }
\renewcommand{\theequation}{A.\arabic{equation}}
\setcounter{equation}{0}

We apply here the effective theory approach discussed in Section 3 to the
original model with a dynamical seesaw mechanism \cite{dhseesaw}.
The transformation properties of the third generation fermions under the
gauge group are shown in Table 1.
The breaking of the gauge group down to the standard model one
leaves a degenerate octet of massive ``colorons'' and
two heavy $U(1)$ gauge bosons. It is assumed that all these gauge
bosons have a mass $M \sim {\cal O}(50$ TeV).

The coefficients of the four-fermion operators are given by
\be
z_{AB} = \frac{2}{\pi} \left( \frac{N_c^2 - 1}{2 N_c}\kappa +
Y_A Y_B \kappa_1 + X_A X_B \kappa_{B-L} \right) ~,
\label{zab}
\ee
where $Y$ are the $U(1)_1$ charges while $X$ are the $U(1)_{B-L}$ charges,
shown in Table 1, and $\kappa, \kappa_1, \kappa_{B-L}$ are the 
$SU(3)_1 \times U(1)_1 \times U(1)_{B-L}$ coupling constants, defined as
the gauge couplings squared divided by $8\pi$.

\begin{table}[htbp]
\centering
\begin{tabular}{|c||c|c|c|c|c|c|}\hline
& $SU(3)_1$ & $SU(3)_2$ & $SU(2)_W$ & $U(1)_1$ & $U(1)_2$ 
& $U(1)_{B-L}$ \vbr
\\\hline \hline
$\psi_L$ & {\bf 3} & {\bf 1} & {\bf 2} & $1/3$  & $0$ & $1/3$ \vbr\\ \hline
$t_R$ & {\bf 3} & {\bf 1} & {\bf 1} & $4/3$  & $0$ & $-1/3 < x < 0$ \vbr\\
\hline
$b_R$ & {\bf 3} & {\bf 1} & {\bf 1} & $-2/3$  & $0$ & $1/3$ \vbr\\ \hline
\hline
$l_L$ & {\bf 1} & {\bf 1} & {\bf 2} & $-1$   & $0$ & $-1$ \vbr\\ \hline
$\tau_R$ & {\bf 1} & {\bf 1} & {\bf 1} & $-2$ & $0$ & $-1$ \vbr\\ \hline
$\nu^{\tau}_R$ & {\bf 1} & {\bf 1} & {\bf 1} & $0$ & $0$ & $-1$ \vbr\\ \hline
\hline
$\chi_L$ & {\bf 3} & {\bf 1} & {\bf 1} & $4/3$ & $0$ & $-1/3 < x < 0$
\vbr\\ \hline
$\chi_R$ & {\bf 3} & {\bf 1} & {\bf 1} & $4/3$ & $0$ & $1/3$
\vbr\\ \hline
\end{tabular}
\caption{Third-generation and $\chi$ fermion representations}
\end{table}

The charge assignment implies $M_{\chi t}^2 < M_{\chi \chi}^2 < 
M_{t t}^2 < M_{\chi b}^2$ and $M_{t \chi}^2 < M_{t b}^2 < M_{\chi b}^2$.
The scalars having $b_R$ as constituent do not acquire VEVs provided
$M_{t b}^2 > 0$, which gives:
\be
- 2 \kappa_1 + \kappa_{B-L} < 12 \left(\frac{3\pi}{8} - \kappa\right) ~.
\label{nobr}
\ee
The vacuum alignment condition $M_{t \chi}^2 < 0 < M_{t t}^2$ requires
\be
4 \kappa_1 + 3x \kappa_{B-L} <
12 \left(\frac{3\pi}{8} - \kappa\right) <
4 \kappa_1 + \kappa_{B-L} ~.
\label{signs}
\ee
Finally, the restriction $M_{\chi t}^2 \gae M_{t \chi}^2$ imposed by 
the minimization condition (\ref{restr}) gives
\be
12 \kappa_1 \lae \left(1 - 9 x^2\right) \kappa_{B-L} ~.
\label{restr1}
\ee
The range of the $U(1)_{B-L}$ charge of $t_R$ and $\chi_L$,  $-1/3 < x < 0$, allows
the  conditions (\ref{nobr}), (\ref{signs}) and (\ref{restr1})
to be simultaneously satisfied.

The relation between the coefficients of the four-fermion operators
and the fermion charges leads to relations among the six $M_{AB}^2$ parameters 
from the effective potential. These relations are simplified by observing that 
the non-Abelian coupling constant $\kappa$ 
is assumed to be larger than the $U(1)$ coupling constants, which implies 
the criticality condition: 
\be
\kappa = \frac{3\pi}{8} + {\cal O}(\kappa_1, \kappa_{B-L}) ~.
\ee
To first order in $\kappa_1/\kappa$ and $\kappa_{B-L}/\kappa$ one can write down 
three sum rules:
\bear
M_{t t}^2 - M_{\chi \chi}^2 & \approx & M_{\chi b}^2 - M_{t t}^2 \approx 
2\left(M_{t b}^2 - M_{t \chi}^2 \right)
\nonumber \\ [2mm]
M_{\chi \chi}^2 - M_{\chi t}^2 & \approx & -3x \left( M_{t t}^2 - M_{t \chi}^2 \right) ~.
\eear
A consequence of the second sum rule is $M_{\chi b} > |M_{t \chi}|$, so that 
the $H^\pm_{tb}$ charged scalar discussed in Section 3.4 is heavier than 
$m_{t\chi}/\epsilon$. Therefore, in addition to $h^0$, the only composite scalars
 that may be lighter than $m_{t\chi}/\epsilon$ are the neutral states 
$A^0_{\chi\chi}$, $A^0_{\chi t}$, $H^0_{\chi t}$, and $\phi_{bb}$.

The composite scalar spectrum is a function of the following parameters:
$\kappa$, $\kappa_1$, $\kappa_{B-L}$, $x$, $\epsilon$ and $\ln (M/m_\chi)$.
For example, the lightest Higgs boson has a mass:
\be
M^2_{h^0} = 4 m_{t\chi}^2 \;
\frac{(1 - 3x) \kappa_{B-L} - 12 \kappa_1}{9\pi \left[ 1 - 3\pi/(8\kappa) \right]
+ 3(1 - x) \kappa_{B-L} - 4 \kappa_1} \, 
\left[1 + {\cal O}(\kappa_1, \kappa_{B-L}, b^2,\epsilon^2)\right] ~.
\ee
In this model, the Higgs boson would have a mass of order 100 GeV only if 
the ratio $\kappa_1/\kappa_{B-L}$ is smaller than $(1-3x)/12$ by no more than a few 
percent.

\section*{Appendix B: Equivalence between the Gap Equations and the
Stationarity Conditions for the Effective Potential}
\renewcommand{\theequation}{B.\arabic{equation}}
\setcounter{equation}{0}

In this Appendix we show that the set of coupled gap equations used in 
ref.~\cite{dhseesaw} is identical [in the large $N_c$ limit and for large 
$\ln (M^2/\overline{M}^2)$] with the stationarity conditions for the
effective potential (see Section 3.2).

The four-fermion operators discussed in Section 3 [see eqs.~(\ref{ops1})
and (\ref{zbr})]
lead to a dynamical mass matrix for the $t$ and $\chi$ quarks,
given in the weak eigenstate basis by
\be
{\cal L} = - \left( \overline{t}_L \ , \ \overline{\chi}_L \right)
\left( \ba{rl} m_{t t} & m_{t \chi} \\
    m_{\chi t} & m_{\chi \chi} \ea \right)
\left( \ba{c} t_R \\ \chi_R \ea \right) + {\rm h.c.} ~,
\label{massterm}
\ee
with all the elements real (this can be ensured by a phase
redefinition of the fields). The top and $\chi$ masses are the eigenvalues of
this matrix,
\bear
m_{t,\chi}^2 & = & \frac{1}{2} \left[ m_{\chi\chi}^2 + m_{t t}^2 +
m_{\chi t}^2 + m_{t \chi}^2 \right.
 \nonumber \\ [3mm]
& & \mp \left.
\sqrt{
\left(m_{\chi\chi}^2 + m_{t t}^2 + m_{\chi t}^2 + m_{t \chi}^2\right)^2
- 4 \left( m_{\chi\chi}m_{t t} - m_{t \chi} m_{\chi t}\right)^2
} \,
\right]
\label{mtmc}
\eear
while the mass eigenstates are given by
\bear
\left( \ba{c} t_L^\prime \\ \chi_L^\prime \ea \right) & = &
\left( \ba{rl} c_L & - s_L \\ s_L & c_L \ea \right)
\left( \ba{c} t_L \\ \chi_L \ea \right) ~,
\nonumber \\ [3mm]
\left( \ba{c} t_R^\prime \\ \chi_R^\prime \ea \right) & = &
\left( \ba{rl} c_R & s_R \\ - s_R & c_R \ea \right)
\left( \ba{c} - t_R \\ \chi_R \ea \right) ~,
\label{massterm2}
\eear
where
\be
c_L,s_L = \frac{1}{\sqrt{2}} \left[ 1 \pm
\frac{ m_{\chi\chi}^2 - m_{t t}^2 + m_{\chi t}^2 - m_{t \chi}^2 }
{m_{\chi}^2 - m_{t}^2} \right]^{\!\! 1/2} ~,
\label{mix}
\ee
and $c_R,s_R$ are obtained by substituting
$m_{t \chi} \leftrightarrow m_{\chi t}$ in the above expressions for
$c_L,s_L$.

The one-loop gap equations can be easily computed by keeping the weak
eigenstates in the external lines, and the $\chi$ and $t$ mass eigenstates
running in the loop (see Fig.~2), and are given by:
\bear
m_{t t} & = & z_{tt} \left\{ - c_L c_R m_t \left[ 1 -
\frac{m^2_t}{M^2}\ln\left( \frac{M^2}{m^2_t} \right)\right]
+ s_L s_R m_\chi \left[ 1 -
\frac{m^2_\chi}{M^2}\ln\left( \frac{M^2}{m^2_\chi} \right)\right]\right\}
  \nonumber \\ [3mm]
m_{\chi\chi} & = & \mu_{\chi\chi} + z_{\chi\chi}
\left\{ - s_L s_R m_t \left[ 1 -
\frac{m^2_t}{M^2}\ln\left( \frac{M^2}{m^2_t} \right)\right]
+ c_L c_R m_\chi \left[ 1 -
\frac{m^2_\chi}{M^2}\ln\left( \frac{M^2}{m^2_\chi} \right)\right]\right\}
   \nonumber \\ [3mm]
m_{t \chi} & = & z_{t \chi} \left\{ c_L s_R m_t \left[ 1 -
\frac{m^2_t}{M^2}\ln\left( \frac{M^2}{m^2_t} \right)\right]
+ s_L c_R m_\chi \left[ 1 -
\frac{m^2_\chi}{M^2}\ln\left( \frac{M^2}{m^2_\chi} \right)\right]\right\}
  \nonumber \\ [3mm]
m_{\chi t} & = & \mu_{\chi t} + z_{\chi t} \left\{ s_L c_R m_t \left[1 -
\frac{m^2_t}{M^2}\ln\left( \frac{M^2}{m^2_t} \right)\right]
+ c_L s_R m_\chi \left[ 1 -
\frac{m^2_\chi}{M^2}\ln\left( \frac{M^2}{m^2_\chi} \right)\right]\right\}
\label{set}
\eear

\begin{figure}[htbp]
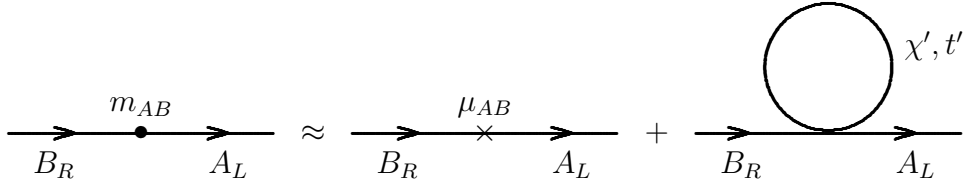

\centering
\hspace*{1.5cm}\parbox{5.5in}{
\beginpicture
\setcoordinatesystem units <\tdim,\tdim>
\stpltsmbl
\moose{-190 0}{-165 0}{-140 0} 
\moose{-140 0}{-107 0}{-90 0} 
\put {$m_{AB}$} at -140 10
\put {$B_R$} at -173 -12
\put {$A_L$} at -107 -12
\put {$\bullet$} at -140 0
\put {$\approx$} at -75 0
\moose{-60 0}{-35 0}{-10 0} 
\moose{-10 0}{23 0}{40 0} 
\put {$\mu_{AB}$} at -10 10
\put {$B_R$} at -43 -12
\put {$A_L$} at 23 -12
\put {$\times$} at -10 0
\put {+} at 55 0
\circulararc 360 degrees from 120 1 center at 120 25
\put {$\chi^\prime, t^\prime$} at 160 33
\moose{70 0}{90 0}{120 0} 
\moose{120 0}{153 0}{170 0} 
\put {$B_R$} at 87 -12
\put {$A_L$} at 153 -12
\endpicture }

\caption{ Coupled gap equations ($A,B = t, \chi$). } 
\end{figure}

Using the relation between the mass matrices in the two basis, namely
\be
\left( \ba{rl} m_{t t} & m_{t \chi} \\
    m_{\chi t} & m_{\chi \chi} \ea \right)
= \left(
\ba{rl} - c_L c_R m_t + s_L s_R m_\chi & c_L s_R m_t + s_L c_R m_\chi \\
     s_L c_R m_t + c_L s_R m_\chi & - s_L s_R m_t + c_L c_R m_\chi
\ea \right) ~,
\label{bastrans}
\ee
we can rewrite the gap equations as
\bear
m_{t t} M^2 \left( \frac{1}{z_{t t}} - 1 \right)
& = & - \left[ m_{tt} \left( m_{tt}^2 + m_{t\chi}^2 + m_{\chi t}^2 \right)
   + m_{t\chi}m_{\chi t}m_{\chi\chi} \right]
\ln\left( \frac{M^2}{m^2_\chi} \right)
   \nonumber \\ [3mm]
m_{\chi\chi} M^2 \left( \frac{1}{z_{\chi\chi}} - 1 \right) -
\mu_{\chi\chi} M^2 \frac{1}{z_{\chi\chi}}
& = & - \left[ m_{\chi\chi} \left(m_{t\chi}^2 + m_{\chi t}^2 + m_{\chi\chi}^2
\right)
   + m_{tt}m_{t \chi}m_{\chi t} \right]
\ln\left( \frac{M^2}{m^2_\chi} \right)
  \nonumber \\ [3mm]
m_{t \chi} M^2 \left( \frac{1}{z_{t \chi}} - 1 \right)
& = & - \left[ m_{t\chi} \left( m_{tt}^2 + m_{t\chi}^2 + m_{\chi\chi}^2\right)
   + m_{tt}m_{\chi t}m_{\chi\chi} \right]
\ln\left( \frac{M^2}{m^2_\chi} \right)
  \nonumber \\ [3mm]
m_{\chi t} M^2 \left( \frac{1}{z_{\chi t}} - 1 \right) -
\mu_{\chi t} M^2 \frac{1}{z_{\chi t}}
& = & - \left[ m_{t\chi} \left(m_{tt}^2 + m_{\chi t}^2 + m_{\chi\chi}^2
\right)
   + m_{tt}m_{t \chi}m_{\chi\chi}\right]
\ln\left( \frac{M^2}{m^2_\chi} \right)
  \nonumber \\
\label{set2}
\eear
where we neglected $m_t^3 \ln(m^2_\chi/m^2_t)$
compared with $m_{t,\chi}^3 \ln(M^2/m^2_\chi)$, which is consistent with
the leading log approximation used in eqs.~(\ref{set}).

One can see that the gap equations (\ref{set2})
are identical with the stationarity conditions (\ref{stationarity})
for the effective potential derived in Section 3.


\vfil
\end{document}